\documentclass[aps,showpacs,twocolumn]{revtex4}
\usepackage{amsfonts}
\usepackage{amsmath}
\usepackage{amssymb}
\usepackage{graphicx}
\usepackage{bm}

\input{tcilatex}

\begin{document}

\title{Do thermal diffusion and Dufour coefficients satisfy Onsager's
reciprocity relation?}
\author{Alois W\"{u}rger}
\affiliation{LOMA, Universit\'{e} de Bordeaux \& CNRS, 351 cours de la Lib\'{e}ration,
33405 Talence, France}

\begin{abstract}
It is commonly admitted that in liquids the thermal diffusion and Dufour
coefficients $D_{T}$\ and $D_{F}$ satisfy Onsager's reciprocity. From their
relation to the cross-coefficients of the phenomenological equations, we are
led to the conclusion that this is not the case in general. As illustrative
and physically relevant examples, we discuss micellar solutions and
colloidal suspensions, where $D_{T}$ arises from chemical reactions or
viscous effects but is not related to the Dufour coefficient $D_{F}$. The
situation is less clear for binary molecular mixtures; available
experimental and simulation data do not settle the question whether $D_{T}$\
and $D_{F}$ are reciprocal coefficients.

PACS\ numbers 05.60.Cd; 05.70.Ln; 66.10.C-; 82.70.-y
\end{abstract}

\maketitle

\section{Introduction}

Onsager's theory for irreversible processes provides a formal framework for
non-equilibrium phenomena as diffusion, electrokinetic effects, and heat
conduction. The phenomenological equations relate thermodynamic fluxes to
forces, for example heat flow to a temperature gradient, or diffusion to a
concentration gradient. Intriguing physical properties arise from the
cross-terms, such as thermally driven electric currents. Onsager established
reciprocal laws between conjugate cross-coefficients \cite{Ons31}, thus
completing Thomson's derivation for the thermoelectric effects and showing
why the Seebeck and Peltier coefficients $\Pi $ and $S$ differ merely by a
temperature factor, $\Pi /S=T$ \cite{deG84}.

As another classical example, the Soret and Dufour effects describe the mass
transport in a temperature gradient, and heat flow due to a concentration
gradient. There is, however, a long-standing discussion whether, and
eventually under which conditions, the corresponding \ coefficients obey a
reciprocity relation. Thus it has been pointed out that the reciprocal laws
impose rather strong conditions on the choice of fluxes and forces, which
are not always easily verified \cite{Col60}.

The Soret and Dufour effects are often discussed in terms of Onsager's
phenomenological equations for the heat and particle currents with
cross-coefficients $L_{1Q}^{\prime }$ and $L_{Q1}^{\prime }$. It is then 
\textit{assumed} that the measured thermal diffusion and Dufour coefficients 
$D_{T}$\ and $D_{F}$ correspond to $L_{1Q}^{\prime }$ and $L_{Q1}^{\prime }$%
, such that the reciprocity relation for the latter applies equally well to
the former. This approach has been taken by various authors, when discussing
binary gases \cite{Ras66}, organic liquids \cite{Har13}, molecular isotope
mixtures \cite{Mor80}, premelting solids with colloidal inclusions \cite%
{Pep09}, and far-from-equilibrium systems \cite{Ge14}.

Available experiments do not provide clear evidence for or against this
assumption. Comparing thermal diffusion and Dufour data for gas mixtures,
suggests qualitative agreeement \cite{Ras66}. Regarding liquids, $D_{T}$\
and $D_{F}$ seem to agree well for mixtures of cyclohexane and
carbon-tetrachloride \cite{Ras70,Hor78}, yet significant discrepancies were
reported for benzene-cyclobenzene and other systems \cite{Ras70}. In recent
years the Soret effect of colloidal suspensions has been investigated in
great detail \cite{Wie04,Iac06,Seh14,Wue10}; unfortunately, there is a lack
of corresponding Dufour data. Molecular dynamics simulations show a good
agreement of the thermal diffusion and Dufour coefficients, at least for
simple model systems \cite{Mac86,Pao87,Mil13,Arm14}.

Here we discuss the validity of the reciprocity assumption for $D_{T}$\ and $%
D_{F}$ on the basis of non-equilibrium thermodynamics. We consider regular
systems where Onsager cross-coefficients are identical, e.g., $%
L_{1Q}^{\prime }=L_{Q1}^{\prime }$. Then the title of this paper reduces to
the question whether, and eventually under which conditions, thermal
diffusion is described by $L_{1Q}^{\prime }$. This is formalized in Sects. 2
and 3, where we define $D_{T}$\ and $D_{F}$, and present Onsager's
phenomenological equations. In Sect. 4 we discuss the case where the entropy
production is given by the vector fluxes, i.e., by the heat flow and
particle currents. In Sects. 5 and 6 we add chemical reactions and viscous
effects; the corresponding contributions to $D_{T}$ are evaluated for the
examples of micellar suspensions, polymers, and colloidal particles. The
final Sections\ 7 discuss and summarize our main results.

\section{Thermal diffusion and Dufour coefficients}

We consider a binary system\ with non-uniform temperature and composition.
Closely following \cite{deG84}, we present the linear equations for heat and
particle flows of a binary systems with volume fractions $\phi _{1}$ and $%
\phi _{2}=1-\phi _{1}$.

Then the currents of the two components satisfy $\mathbf{J}_{1}+\mathbf{J}%
_{2}=0$; the first one is defined as 
\begin{equation}
\mathbf{J}_{1}=-D\mathbf{\nabla }\phi _{1}-\phi _{1}\phi _{2}D_{T}\mathbf{%
\nabla }T.  \label{2}
\end{equation}%
Besides gradient diffusion with the coefficient $D$, it comprises thermal
diffusion with coefficient $D_{T}$. In the steady state $\mathbf{J}_{1}=0$,
a finite temperature gradient imposes a non-uniform concentration $\mathbf{%
\nabla }\phi _{1}=-\phi _{1}\phi _{2}S_{T}\mathbf{\nabla }T$, with the Soret
coefficient $S_{T}=D_{T}/D$.

Similarly, the heat flow is driven by both temperature and concentration
gradients, 
\begin{equation}
\mathbf{J}_{Q}^{\prime }=-\lambda \mathbf{\nabla }T-\phi _{1}\hat{\mu}%
_{11}^{\phi }D_{F}\mathbf{\nabla }\phi _{1},  \label{4}
\end{equation}%
where $\lambda $ is the thermal conductivity, $D_{F}$\ the Dufour
coefficient, and $\hat{\mu}_{11}^{\phi }$ the derivative of the chemical
potential \cite{deG84}. The hat indicates volume specific quantities, such
as the molecular chemical potential divided by the molecular volume, $\hat{%
\mu}_{k}=\mu _{k}/v_{k}$. The choice of volume fraction variables $\phi _{k}$
implies that $\mathbf{J}_{1}$ has the dimension of a velocity; one readily
finds that $D_{T}$ has the units m$^{2}$s$^{-1}$K$^{-1}$, whereas $D_{F}$\
has those of a diffusion coefficient, m$^{2}$s$^{-1}$.

The heat flow $\mathbf{J}_{Q}^{\prime }$ comprises two contributions,
ordinary heat diffusion with conductivity $\lambda $, and the Dufour effect
which is driven by a concentration gradient in an anisotropic medium. Note
that $\mathbf{J}_{Q}^{\prime }$ does not account for enthalpy transport due
to the particle current but is defined at $\mathbf{J}_{1}=0$. In the case of
a finite particle current, the total heat flow reads 
\begin{equation}
\mathbf{J}_{Q}=\mathbf{J}_{Q}^{\prime }+\hat{h}_{1}\mathbf{J}_{1}+\hat{h}_{2}%
\mathbf{J}_{2}=\mathbf{J}_{Q}^{\prime }+(\hat{h}_{1}-\hat{h}_{2})\mathbf{J}%
_{1},  \label{5}
\end{equation}%
with the specific enthalpies $\hat{h}_{k}$ of the two components.

In many instances it is assumed that the off-diagonal coefficients $D_{T}$\
and $D_{F}$\ are related through Onsager's reciprocal laws according to 
\begin{equation}
D_{T}\overset{?}{=}\frac{D_{F}}{T}.  \label{6}
\end{equation}%
The temperature factor is due to historical convention, similar to that
between the Seebeck and Peltier coefficients, $S=\Pi /T$. Kinetic theory
confirms this relation to be satisfied in ideal gas mixtures \cite{Cha60},
and there is evidence for its validity in liquid isotope mixtures \cite%
{Mor80}. Little can be said on ordinary binary liquids \cite{Ras70,Hor78}
and complex fluids.

\section{Phenomenological equations}

The above Eqs. (\ref{2}) and (\ref{4}) provide the experimental definition
of the thermal diffusion and Dufour coefficients.\ Closely following Ref. 
\cite{deG84}, we summarize the corresponding theory, that is, Onsager's
linear relations for thermodynamic fluxes and forces. We do not discuss the
regression hypothesis \cite{Ell00} and suppose that the fluxes are linear
functions of the forces; note this assumption is often not justified for
chemical reactions.

\subsection{Entropy production}

Like any thermodynamic function, the entropy is constant in an equilibrium
state. Non-equilibrium phenomena are intimately related to entropy
production. For example, gradient diffusion according to Fick's law $\mathbf{%
J}_{1}=-D\mathbf{\nabla }\phi _{1}$ tends to smear out any \ composition
inhomogeneity and produces entropy at a rate $\sigma \propto D(\mathbf{%
\nabla }\phi _{1})^{2}$. By the same token, a non-uniform temperature
induces a heat flow $\mathbf{J}_{Q}=-\lambda \mathbf{\nabla }T$ from the hot
to the cold and augments the entropy as $\sigma =\lambda (\mathbf{\nabla }%
T/T)^{2}$. Similar relations arise for chemical reactions and for viscous
flow.

In the case of an initial perturbation, the system relaxes toward an
equilibrium state ($\mathbf{\nabla }\phi _{1}=0$, $\mathbf{\nabla }T=0$,...)
of constant entropy. On the other hand, if the inhomogeneity is maintained
through continuous heat or matter supply, the system reaches a stationary
non-equlibrium state and produces entropy at a constant rate.

With the mentioned dynamic variables, the rate of entropy production per
unit volume reads as

\begin{equation}
\sigma =\mathbf{J}_{Q}\cdot \mathbf{\nabla }\frac{1}{T}-\sum_{k}\mathbf{J}%
_{k}\cdot \mathbf{\nabla }\frac{\hat{\mu}_{k}}{T}-\sum_{i}\mathcal{J}%
_{i}A_{i}-\frac{\mathbf{\Pi :G}}{T}\mathbf{,}  \label{12}
\end{equation}%
where $\mathbf{J}_{Q}$ is the heat flux, $\mathbf{J}_{k}$ are the volume
currents of the molecular species, $\mathcal{J}_{i}$ are the compositon
changes due to chemical reactions, and $\mathbf{\Pi }$ is the viscous
pressure tensor. The corresponding thermodynamic forces are the gradients of
the inverse temperature and the Planck potential $\hat{\mu}_{k}/T$, the
affinities $A_{i}$, and the symmetrized rate of change of the fluid velocity
field $\mathbf{v(r)}$, with components $G_{mn}=\frac{1}{2}(\partial
_{m}v_{n}+\partial _{n}v_{m})$.\ 

For sufficiently weak deviations from the equilibrium state, Onsager
established linear relations between the fluxes and forces. Because of the
Curie symmetry principle, the phenomenlogical relations do not mix scalar,
vector, and tensor quantities. Thus the coefficient matrix of the
phenomenological equations is block-diagonal and decays in parts that are
characterized by their tensor properties.

\subsection{Vector currents}

The vector quantities $\mathbf{J}_{Q}$ and $\mathbf{J}_{k}$ describe heat
and mass diffusion. For a binary system ($n=2$) they\ form 3 generalized
fluxes which are, however, not linearly independent and can be reduced to 2
independent flows.\ When describing the composition in terms of volume
fractions, the particle currents cancel each other, $\mathbf{J}_{2}=-\mathbf{%
J}_{1}$; eliminating that of the second component one obtains 
\begin{subequations}
\label{15}
\begin{eqnarray}
\mathbf{J}_{1} &=&L_{1Q}\mathbf{\nabla }\frac{1}{T}-L_{11}\mathbf{\nabla }%
\frac{\hat{\mu}_{1}-\hat{\mu}_{2}}{T}\mathbf{,}  \label{15a} \\
\mathbf{J}_{Q} &=&L_{QQ}\mathbf{\nabla }\frac{1}{T}-L_{Q1}\mathbf{\nabla }%
\frac{\hat{\mu}_{1}-\hat{\mu}_{2}}{T}\mathbf{.}  \label{15b}
\end{eqnarray}%
The last term of each equation gives rise to both thermal and concentration
gradients, 
\end{subequations}
\begin{equation}
\mathbf{\nabla }\frac{\hat{\mu}_{k}}{T}=\hat{h}_{k}\mathbf{\nabla }\frac{1}{T%
}+\frac{\mathbf{\nabla }_{T}\hat{\mu}_{k}}{T},  \label{24}
\end{equation}%
where $\hat{h}_{i}$ is the enthalpy and $\mathbf{\nabla }_{T}$ the gradient
at constant temperature. Thus the thermodynamic force $\mathbf{\nabla }_{T}%
\hat{\mu}_{k}$ involves the derivative of the chemical potential with
respect to composition.

In many instances it turns out convenient to regroup all temperature
gradients according to 
\begin{subequations}
\label{16}
\begin{eqnarray}
\mathbf{J}_{1} &=&L_{1Q}^{\prime }\mathbf{\nabla }\frac{1}{T}-L_{11}^{\prime
}\frac{\mathbf{\nabla }_{T}(\hat{\mu}_{1}-\hat{\mu}_{2})}{T}\mathbf{,}
\label{16a} \\
\mathbf{J}_{Q}^{\prime } &=&L_{QQ}^{\prime }\mathbf{\nabla }\frac{1}{T}%
-L_{Q1}^{\prime }\frac{\mathbf{\nabla }_{T}(\hat{\mu}_{1}-\hat{\mu}_{2})}{T}%
\mathbf{.}  \label{16b}
\end{eqnarray}%
Comparison with (\ref{12}) readily provides relations between unprimed and
primed coefficients, e.g. $L_{1Q}^{\prime }=L_{1Q}-L_{11}(\hat{h}_{1}-\hat{h}%
_{2})$.

The heat flow $\mathbf{J}_{Q}^{\prime }$\ is defined such that the entropy
production involves products of conjugate forces and currents; for a binary
system with $\mathbf{J}_{1}+\mathbf{J}_{2}=0$, the contribution of the
vector quantities reads 
\end{subequations}
\begin{equation*}
\mathbf{J}_{Q}^{\prime }\cdot \mathbf{\nabla }\frac{1}{T}-\mathbf{J}%
_{1}\cdot \frac{\mathbf{\nabla }_{T}(\hat{\mu}_{1}-\hat{\mu}_{2})}{T}\mathbf{%
.}
\end{equation*}%
According to (\ref{5}) the primed heat flux accounts for diffusive transport
only, whereas $\mathbf{J}_{Q}$ comprises in addition the enthalpy carried by
the particle current $\mathbf{J}_{1}$. In the steady-state of a closed
system, the latter vanishes and one has $\mathbf{J}_{Q}=\mathbf{J}%
_{Q}^{\prime }$.

\subsection{Scalar and tensor quantities}

Now we turn to the remaining terms of the entropy production rate. That
involving chemical reactions is described by scalar fields, 
\begin{equation}
\mathcal{J}_{i}=-\sum_{j}l_{ij}A_{j}/T,  \label{14}
\end{equation}%
where the coefficients $l_{ij}$ relate the reaction products to the
affinities $A_{i}$.

Finally, the linear relation between the viscous pressure and the velocity
gradient, 
\begin{equation}
\mathbf{\Pi }=\mathbf{-\eta G,}  \label{18}
\end{equation}%
involves the fourth-rank viscosity tensor $\mathbf{\eta }$, which structure
is rather simple in isotropic liquids, yet becomes more complex in liquid
crystals \cite{Mon12}. For compressible fluids, the contraction $\mathbf{\Pi
:G}$ comprises also a scalar term, which is small for most liquids and thus
will discarded.

\subsection{Reciprocal laws}

According to Onsager's reciprocal laws, the coefficient matrices $\mathbf{l}$%
, $\mathbf{L}$, $\mathbf{L}^{\prime }$, and $\mathbf{\eta }$ are symmetric,
and in particular%
\begin{equation}
L_{1Q}^{\prime }=L_{Q1}^{\prime }.
\end{equation}%
In many works on thermal diffusion, both chemical reactions and viscous flow
are discarded from the beginning. Then the different terms in (\ref{16}) are
readily identified with those in (\ref{2}) and (\ref{4}), 
\begin{equation}
\lambda \overset{?}{=}\frac{L_{QQ}^{\prime }}{T^{2}},D\overset{?}{=}\frac{%
\hat{\mu}_{11}^{\phi }L_{11}^{\prime }}{\phi _{2}T},D_{T}\overset{?}{=}\frac{%
L_{1Q}^{\prime }}{\phi _{1}\phi _{2}T^{2}},D_{F}\overset{?}{=}\frac{%
L_{Q1}^{\prime }}{\phi _{1}\phi _{2}T},  \label{20}
\end{equation}%
where $\hat{\mu}_{11}^{\phi }$ is the usual derivative with respect to
composition \cite{deG84}. Since $L_{1Q}^{\prime }$ and $L_{Q1}^{\prime }$
are reciprocal coefficients, these relations confirm (\ref{6}) for the
thermal diffusion and Dufour coefficient. \ 

The above decomposition of the linear relations according to their tensorial
properties does not imply, however, that the underlying physical phenomena
are decoupled. Whether each of the fluxes (\ref{16}), (\ref{14}), and (\ref%
{18}) can be treated independently from the others, cannot be determined on
formal grounds, but has to be inferred from the physical properties of the
system under consideration.

In the following we evaluate $D_{T}$ for different models and determine in
each case whether or not Eq. (\ref{20}) is satisfied.

\section{Diffusion}

Here we consider the case of a binary system where both chemical reactions
and viscous effects are absent. Then the entropy production and
phenomenological relations reduce to the vector quantities $\mathbf{J}_{Q}$, 
$\mathbf{J}_{1}$, and $\mathbf{J}_{2}$, implying that $D_{T}$ and $D_{F}$
are reciprocal coefficients according to (\ref{6}). Still, the remaining
three independent coefficients have to be determined from physical
considerations. The general theory parallels Chapt.\ XI \S 8 of Ref. \cite%
{deG84}; the notation with volume fractions and examples are developped in
Ref. \cite{Wue14}.

\subsection{Vector fluxes}

It turns out instructive to eliminate the heat current $\mathbf{J}_{Q}$,
contrary to (\ref{16}) where we eliminated the current of the second
component $\mathbf{J}_{2}$. Then one obtains the particle fluxes as linear
functions of the thermodynamic forces $\mathbf{\nabla }(\hat{\mu}_{k}/T)$, 
\begin{equation}
\mathbf{J}_{1}=-\phi _{1}\phi _{2}T\left( B_{1}\mathbf{\nabla }\frac{\hat{\mu%
}_{1}}{T}-B_{2}\mathbf{\nabla }\frac{\hat{\mu}_{2}}{T}\right) ,  \label{22}
\end{equation}%
where the coefficients $B_{i}$ depend on composition and on temperature. (As
compared to the notation in \cite{Wue14}, a factor $\phi _{1}\phi _{2}T$ has
been introduced for convenience.)

Note that the currents $\mathbf{J}_{k}$ have the dimension of a velocity.
Our choice of volume fraction variables $\phi _{i}$ is motivated by the
volume conservation in incompressible liquids, which results in the relation 
$\mathbf{J}_{1}+\mathbf{J}_{2}=0$. For gases, where momentum is conserved,
one would prefer to take mass fractions.

Eq. (\ref{22}) implies that diffusion and thermal diffusion are given by the
gradient of the Planck potentials $\hat{\mu}_{k}/T$. Spelling out the
gradients,%
\begin{equation}
\mathbf{\nabla }\frac{\hat{\mu}_{k}}{T}=-\frac{\hat{h}_{k}}{T^{2}}\mathbf{%
\nabla }T+\frac{\hat{\mu}_{kk}^{\phi }}{T}\mathbf{\nabla }\phi _{k},
\label{19}
\end{equation}%
recollecting the terms in $\mathbf{J}_{1}$ and comparing with (\ref{2}) one
obtains the diffusion coefficient 
\begin{subequations}
\label{26}
\begin{equation}
D=\left( \phi _{1}B_{1}+\phi _{2}B_{2}\right) \phi _{1}\hat{\mu}_{11}^{\phi
}.  \label{26a}
\end{equation}%
Similarly one finds for the thermal diffusion and Dufour coefficients \cite%
{deG84} 
\begin{equation}
D_{T}=\frac{\hat{h}_{2}B_{2}-\hat{h}_{1}B_{1}}{T}=\frac{D_{F}}{T}.
\label{26b}
\end{equation}

Eqs. (\ref{16}) and (\ref{26}) provide formally equivalent expressions for
the particle current (\ref{2}): The former depends on two unknowns $%
L_{11}^{\prime }$ and $L_{1Q}^{\prime }$, and the latter on $B_{1}$ and $%
B_{2}$. These coefficients can not be derived from equilibrium properties;
they have to be taken from experiment or molecular dynamics simulations, or
inferred from models for the molecular mobility.

On the other hand, the equilibrium quantities appearing in (\ref{26}), that
is, the thermodynamic factor $\phi _{1}\hat{\mu}_{11}^{\phi }$ and the
specific enthalpies $\hat{h}_{i}$, can be calculated from first principles 
\cite{Hel60}, or can be taken from numerical simulations \cite{Mil13} or
experiment \cite{Wue14}.

\subsection{Comparison to experiment}

Thermal diffusion is usually discussed in terms of (\ref{16a}). When
comparing to experiments, however, Eqs. (\ref{26}) turn out to be a more
promising starting point. Two mobilities $B_{k}$ appear in both $D$ and $%
D_{T}$ and thus provide a strong relation between thermal diffusion and
diffusion data. When taking the parameters $B_{k}$ as constants, they can be
determined from the tracer diffusion coefficients of the two species, 
\end{subequations}
\begin{equation}
\begin{array}{ccc}
B_{2}=D(\phi _{1})/\phi _{1}\hat{\mu}_{11}^{\phi } & \text{for} & \phi
_{1}\rightarrow 0, \\ 
B_{1}=D(\phi _{1})/\phi _{1}\hat{\mu}_{11}^{\phi } & \text{for} & \phi
_{1}\rightarrow 1.%
\end{array}
\label{25}
\end{equation}%
A slightly different scheme has been used for Soret data of several
molecular mixtures such as benzene-cyclohexane \cite{Wue14}. Together with
measured values for the partial enthalpies $\hat{h}_{2}$, this allows
comparison with the thermal diffusion coefficient $D_{T}$.

A particularly interesting situation arises for isotope mixtures. Molecular
isotopes show similar thermodynamics, yet differ in dynamical properties
such as the attempt frequency of activated jumps. Thus any difference in the
mobilities $B_{k}$ can be related to a specific model. Such approaches have
been developped for mass effects. For mixtures such as CCl$_{4}$-CBr$_{4}$
the $B_{k}$ have been expressed through the activation free energy \cite%
{Mor80,Sax62}. The isotope effect observed upon deuteration in
benzene-cyclohexane mixtures, has been related to the molecular collision
rates \cite{Vil11}.

\begin{figure}[tbp]
\includegraphics[width=\columnwidth]{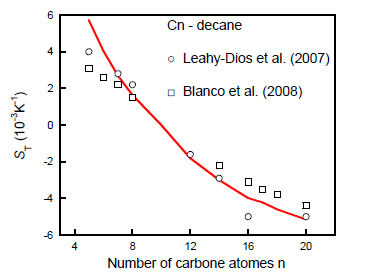}
\caption{Soret data for mixtures of normal alkanes C$_{10}-$C$_{n}$, where
the number $n$ of carbon atoms of the second component varies from 5 to 20.
The experimental data are from Leahy-Dios and Firoozabaadi \protect\cite%
{Lea07} (squares) and Blanco et al. \protect\cite{Bla08} (circles). The
theoretical curve is calculated from (\protect\ref{70}) as discussed in the
main text.}
\label{fig1}
\end{figure}

\subsection{Alkane mixtures}

Thermal diffusion behavior according to (\ref{26}) is expected in mixtures
of similar molecules, such as short alkanes. Indeed, small molecules induce
weak hydrodynamic flow, which is rapidly superseded by the molecular
diffusion. Then the Soret coefficient $S_{T}=D_{T}/D$ depends on the partial
enthalpies $\hat{h}_{k}$, the chemical potentials $\hat{\mu}_{k}$, and the
molecular mobilities $B_{k}$.

Assumung that the latter are identical, independently of the molecular
weight, we thus have \cite{Wue14} 
\begin{equation}
S_{T}=\frac{1}{T}\frac{\hat{h}_{2}-\hat{h}_{1}}{\phi _{1}\hat{\mu}_{11}}\ \
\ \ \ \ \ \ \ (B_{1}=B_{2}).  \label{70}
\end{equation}%
Refining early work by Haase \cite{Haa49}, similar relations have been
discussed by several authors \cite{Kem89,Shu98,Esl09}. In Fig. 1 we compare
this expression with Soret data for equimolar alkane mixtures, which are
taken from Refs. \cite{Lea07,Bla08}.

The theoretical curve has been calculated from (\ref{70}), with the
thermodynamic factor equal to unity such that $\phi _{1}\hat{\mu}%
_{11}=k_{B}T/(\phi _{1}v_{2}+\phi _{2}v_{1})$, and the specific enthalpy $%
\hat{h}_{n}=h_{n}/v_{n}$ given by 
\begin{subequations}
\label{72}
\begin{eqnarray}
h_{n} &=&-(n+0.65)\times 4.64\frac{\text{kJ}}{\text{mol}},  \label{72a} \\
v_{n} &=&(n+2.02)\times 16.35\frac{\text{cm}^{3}}{\text{mol}}.  \label{72b}
\end{eqnarray}%
These simple laws perfectly fit the measured vaporization enthalpy \cite%
{Chi03} and molecular volume $v_{n}$. The theoretical curve in Fig. 1
strongly depends on the off-set parameters $0.65$ and $2.02$, as discussed
in the Appendix.

The particularly simple fit arises since the Soret coefficient $%
S_{T}=D_{T}/D $ depends on the ratio $B_{1}/B_{2}$ only (which, moreover,
has been put to unity.) A more complex situation occurs when considering
thermal diffusion and diffusion data separately.\ The many data on the
composition dependence of $D_{T}$ and $D$ \cite{Mad10,deM12} should
unambiguously determine the mobilities $B_{k}$ and verify whether the
assumption $B_{1}=B_{2}$ is justified. On the other hand, if the data for $%
D_{T}$ and $D$ turned out not to be compatible with (\ref{26}), this would
suggest that $D_{T}$ cannot be explained in terms of the vector fluxes (\ref%
{15}) but depends on other phenomena such as viscous stress.

\subsection{Comparison with heat conductivity}

So far we discussed the relation between diffusion and thermal diffusion
coefficients in terms of the quantities $B_{k}$. As shown by de Groot and
Mazur, a more general formulation of the heat and particle flows relates the
coefficients $D$ and $D_{T}$\ to the heat conductivity $\lambda $. To this
purpose, we note that the thermodynamic forces on component $k$ can be
written as $\sum_{l}a_{kl}\mathbf{\nabla (}\hat{\mu}_{l}/T)$, with a
mobility matrix that is symmetric ($a_{12}=a_{21}$) and positive definite ($%
a_{11}a_{22}\geq a_{21}^{2}$).

The $B_{k}$ appearing in (\ref{22}) depend on these mobilities through 
\end{subequations}
\begin{equation}
B_{1}=\frac{\phi _{2}a_{11}-\phi _{1}a_{12}}{\phi _{1}\phi _{2}T},\ \ \ \
B_{2}=\frac{\phi _{1}a_{22}-\phi _{2}a_{21}}{\phi _{1}\phi _{2}T}.
\label{21}
\end{equation}%
Insertion in (\ref{26}) provides the diffusion and thermal diffusion
coefficients in terms of the $a_{ij}$. Similarly, the heat conductivity is
given by the mobilities according to \cite{deG84}%
\begin{equation}
\lambda =\frac{a_{11}\hat{h}_{1}^{2}+2a_{12}\hat{h}_{1}\hat{h}_{2}+a_{22}%
\hat{h}_{2}^{2}}{T^{2}}.  \label{23}
\end{equation}%
Thus the three transport coefficients $D$, $D_{T}$, and $\lambda $\ are
expressed through three paramaters $a_{11}$, $a_{22}$, and $a_{12}$. These
quantities vary with composition, such that a set of experimental data for $%
D(\phi _{1})$, $D_{T}(\phi _{1})$, and $\lambda (\phi _{1})$ determines the
mobility matrix $a_{kl}(\phi _{1})$.

Here one should remind that (\ref{26}) and (\ref{23}) rely on the assumption
that vector fluxes and forces entirely determined the heat and particle
currents. Thus in the first place, these relations provide a criterion for
the validity of this assumption. As a crude estimate, we replace the
mobilities by $a_{ij}\sim a$, neglect composition factors $\phi _{i}\sim 1$,
and thus obtain $D\sim ak_{B}/v$ and $\lambda \sim a(H/v)^{2}/T^{2}$, \
where $H$ and $v$\ are the molar enthalpy and volume. Eliminating $a$ leads
to $\lambda \sim H^{2}D/vk_{B}T^{2}$; inserting typical values of $H,D,v$,
as measured e.g. for benzene, one finds a thermal conductivity $\lambda \sim
0.03$ Wm$^{-1}$K$^{-1}$; which is about five times smaller than the
experimental value.\ 

This estimate suggests that it could be instructive to fit experimental or
simulation data for $D$, $D_{T}$, and $\lambda $\ with Eqs. (\ref{26}) and (%
\ref{23}). The elements of the mobility matrix $a_{kl}$ do, however, not
necessarily provide a good starting point; thus according to (\ref{21}) and (%
\ref{25}), constant diagonal elements $a_{kk}$ would result in a diverging
diffusion coefficient $D$. In view of (\ref{25}) one would rather prefer to
replace the $a_{kl}$ with well-behaved quantities; as a possible choice we
note $B_{1}$, $B_{2}$, and 
\begin{equation*}
B_{3}=\frac{a_{12}}{\phi _{1}\phi _{2}T}.
\end{equation*}%
Then the expressions in (\ref{26}) are completed by the thermal conductivity 
\begin{equation}
\lambda =\frac{\phi _{1}\hat{h}_{1}^{2}B_{1}+\phi _{2}\hat{h}%
_{2}^{2}B_{2}+(\phi _{1}\hat{h}_{1}+\phi _{2}\hat{h}_{2})^{2}B_{3}}{T}.
\label{27}
\end{equation}%
The positivity condition for the mobility matrix imposes 
\begin{equation*}
B_{1}B_{2}+(\phi _{2}B_{1}+\phi _{1}B_{2})B_{3}\geq 0.
\end{equation*}%
We recall that the $B_{k}$ depend on composition. The relations (\ref{26})
and (\ref{27}) suggest that, in a simple model, these quantities could be
taken as constants.

\section{Chemical reactions}

Here we discuss how chemical reactions modify heat and mass flow. In
physical terms, it is clear they affect the local composition and thus
induce diffusion currents; inversely, thermal diffusion creates a
non-uniform composition which in turn perturbs the chemical equilibrium and
thus provokes reactions.

Because of the Curie principle, the phenomenological equations (\ref{16})
and (\ref{14}) do not contain cross terms between the scalar and vector
quantities, and thus do not mix the reaction kinetics and diffusion. This
does not imply, however, that scalar and vector fluxes are independent of
each other, nor that diffusion is simply determined by the matrix $\mathbf{L}
$.

\subsection{Reaction-diffusion coupling}

Consider the case of a single chemical reaction between two components. Then
the entropy production (\ref{12}) comprises the scalar fluxes $\mathcal{J}%
_{i}=\dot{\phi}_{i}$. The reaction kinetics obey the rate equations 
\begin{equation}
\dot{\phi}_{1}=\gamma \phi _{2}-\Gamma \phi _{1}=-\dot{\phi}_{2},  \label{32}
\end{equation}%
where the point indicates time derivatives and the transitions $%
1\leftrightarrow 2$ occur with rates $\gamma $ and $\Gamma $. According to
the principle of detailed balanced, their ratio $\gamma /\Gamma
=e^{-G/k_{B}T}$ is determined by the free enthalpy difference $G$ of the two
states.

In addition to the reaction velocity $\dot{\phi}_{k}$, the total change of
the volume fraction with time comprises the divergence of the diffusion
current $\mathbf{J}_{k}$,%
\begin{equation}
\frac{d}{dt}\phi _{k}=\dot{\phi}_{k}+\mathbf{\nabla \cdot J}_{k},  \label{36}
\end{equation}%
where the last term corresponds to a source or sink for the considered
species. Eq. (\ref{36}) provides a coupling between the scalar and vector
fluxes $\dot{\phi}_{k}$ and $\mathbf{J}_{k}$, and thus induces a relation
between the a priori independent phenomenological equations (\ref{16}) and (%
\ref{14}).

Here we consider the steady state where $d\phi _{k}/dt=0$. The resulting
equation 
\begin{equation}
\dot{\phi}_{k}+\mathbf{\nabla \cdot J}_{k}=0  \label{38}
\end{equation}%
does not imply the arrest of the reaction nor that the currents $\mathbf{J}%
_{k}$ vanish.\ It simply requires that a local creation of molecules ($\dot{%
\phi}_{1}>0$) is balanced by a net outgoing particle flow, and annihilation (%
$\dot{\phi}_{1}<0$) by an incoming flow. In general the solution of (\ref{38}%
) cannot be given in closed form, especially if the rate ratio $\gamma
/\Gamma =e^{-G/k_{B}T}$ depends on temperature.\ 

Thermal conductivity of a binary system with constant rates has been studied
in detail by de Groot and Mazur in Chapter XI \S 8 of Ref. \cite{deG84}. As
a main result these authors find that the diffusivities are not simply given
by the matrix $\mathbf{L}$, but depend on the reaction parameters; moreover,
the explicit result for the thermal conductivity shows an intricate spatical
variation. This leads to the conclusion that Eq. (\ref{20}) is not valid in
the presence of chemical reactions.

\begin{figure}[tbp]
\includegraphics[width=\columnwidth]{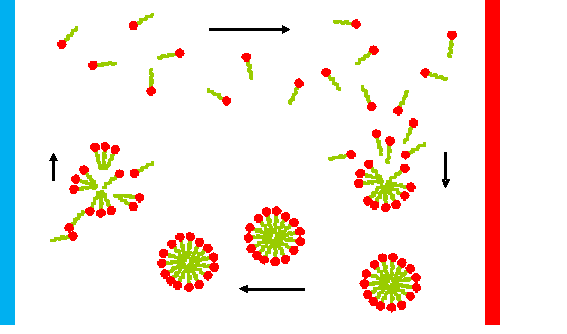}
\caption{Steady-state currents resulting from the equilibrium between
micellar and molecular states in a surfactant solution. For $H>0$, micelle
formation is favored at high temperatures, and the dissolved molecular state
at lower temperatures. This temperature dependent equilibrium feeds
stationary micelle and molecular currents $J_{1}$ and $J_{2}$. For $H<0$ all
arrows point in the opposite direction. }
\end{figure}

\subsection{Micelle kinetics}

As an instructive example, we consider a solution of tensioactive molecules
that partly aggregate to micelles. The micellar and molecular states occupy
volume fractions $\phi _{1}$ and $\phi _{2}$, with $\phi _{1}+\phi _{2}\ll 1$%
. For the sake of simplicity we suppose that the cross-coefficients in the
vector fluxes are small and thus put 
\begin{subequations}
\label{35}
\begin{equation}
L_{kQ}^{\prime }=0=L_{Qk}^{\prime }.
\end{equation}%
Then the particle currents take the form $\mathbf{J}_{k}=-D_{k}\mathbf{%
\nabla }\phi _{k}$, where%
\begin{equation}
D_{k}=\frac{L_{kk}^{\prime }\phi _{k}\hat{\mu}_{kk}^{\phi }}{\phi _{1}\phi
_{2}T},\ \ \ \ \ D_{T}=0
\end{equation}%
implies the absence of thermal diffusion.

Now we take the kinetics of micelle formation into account \cite{Kaa09}.\
Micellar aggregation of $N$ molecules and dissolution through the inverse
process, as described by the rate equation (\ref{32}). This
\textquotedblleft chemical reaction\textquotedblright\ occurs on a time
scale of nanoseconds, and thus is much faster than diffusion over
macroscopic lengths $L$, which occurs on a time scale $L^{2}/D$ that by far
exceeds seconds. This means that the second term in (\ref{38}) is small.

The steady-state of a homogeneous system is determined by the equilibrium of
the chemical reaction ($\dot{\phi}_{1}=0$) which reads as $\phi _{1}/\phi
_{2}=\gamma /\Gamma =e^{-G/k_{B}T}$. Yet here this ratio is not constant in
space but varies because of the non-uniform temperature $T(\mathbf{r)}$.
Expanding the rate equation (\ref{32}) to linear order in the coordinates $%
\mathbf{r}$, we have 
\end{subequations}
\begin{equation}
\gamma \phi _{2}-\Gamma \phi _{1}+\mathbf{r\cdot \nabla }\left( \gamma \phi
_{2}-\Gamma \phi _{1}\right) +\mathbf{\nabla \cdot J}_{1}=0.
\end{equation}%
The first term describes the equilibrium state at $\mathbf{r}=0$. Since
diffusion is much slower than the chemical reaction, $D/L^{2}\ll \gamma $,
the last term is negligible. Evaluating the gradient of the remainder we
find 
\begin{equation}
0=\Gamma \phi _{1}\frac{H}{k_{B}T^{2}}\mathbf{\nabla }T+\gamma \mathbf{%
\nabla }\phi _{2}-\Gamma \mathbf{\nabla }\phi _{1},  \label{37}
\end{equation}%
where we have inserted the Gibbs-Helmholtz relation 
\begin{equation*}
\frac{d}{dT}\frac{G}{k_{B}T}=-\frac{H}{k_{B}T^{2}}
\end{equation*}%
with the enthalpy $H$.

Thus chemical equilibrium in the presence of a non-uniform temperature
gradient imposes gradients of the micellar and molecular volume fractions,
which in turn drive the diffusion currents $\mathbf{J}_{k}=-D_{k}\mathbf{%
\nabla }\phi _{k}$. Inserting $\mathbf{J}_{1}+\mathbf{J}_{2}=0$ in (\ref{37}%
) we find the steady-state composition gradients 
\begin{equation}
\mathbf{\nabla }\phi _{1}^{\text{st}}=\phi _{1}\frac{D_{2}\Gamma }{%
D_{1}\gamma +D_{2}\Gamma }\frac{H}{k_{B}T^{2}}\mathbf{\nabla }T  \label{39}
\end{equation}%
and $\mathbf{\nabla }\phi _{2}^{\text{st}}=-(D_{1}/D_{2})\mathbf{\nabla }%
\phi _{1}^{\text{st}}.$ This implies a finite stationary current of micelles 
\begin{equation}
\mathbf{J}_{1}^{\text{st}}=-\phi _{1}\frac{D_{1}D_{2}\Gamma }{D_{1}\gamma
+D_{2}\Gamma }\frac{H}{k_{B}T^{2}}\mathbf{\nabla }T=-\phi _{1}D_{T}\mathbf{%
\nabla }T,  \label{40}
\end{equation}%
and the opposite flow of the molecular state. The second equality defines a
thermal diffusion coefficient $D_{T}$ for the first component. Its
dependence on the rates $\gamma $\ and\ $\Gamma $, and on the micellar
enthalpy $H$, indicates that this current is driven by the chemical
reaction. The micelle and molecular currents $\mathbf{J}_{1}^{\text{st}}$
and $\mathbf{J}_{2}^{\text{st}}$ are shown in Fig. 2 and related to the
gradient of the chemical equilibrium condition (\ref{37}).

Finally we define the Soret coefficient of the solution through the gradient
of the total surfactant content $\phi ^{\text{st}}=\phi _{1}^{\text{st}%
}+\phi _{2}^{\text{st}}$, 
\begin{equation*}
\mathbf{\nabla }\phi ^{\text{st}}+\phi ^{\text{st}}S_{T}\mathbf{\nabla }T=0.
\end{equation*}%
Rearranging the above expressions one finds 
\begin{equation}
S_{T}=\frac{D_{1}-D_{2}}{D_{1}\gamma +D_{2}\Gamma }\frac{\gamma \Gamma }{%
\gamma +\Gamma }\frac{H}{k_{B}T^{2}}\mathbf{.}
\end{equation}%
This contribution could be relevant for thermophoresis experiments on SDS
solutions\ \cite{Vig10}.

As a summary of this Section, the non-uniform equilibrium condition (\ref{37}%
) imposes the steady-state composition gradient $\mathbf{\nabla }\phi _{1}^{%
\text{st}}$, which in turn, induces the thermally driven micellar current $%
\mathbf{J}_{1}^{\text{st}}$.\ Thus we have $D_{T}\neq 0$ in spite of $%
L_{kQ}^{\prime }=0$, which clearly invalidates Eq. (\ref{20}).

\section{Viscous effects $-$ thermophoresis}

Now we turn to the case where the entropy production (\ref{12}) comprises a
viscous term. We repeat that the phenomenological equations are
block-diagonal, and thus do not mix viscous and particle fluxes\textbf{\ }$%
\Pi $ and $\mathbf{J}_{k}$. Yet there is ample experimental evidence that
thermally driven motion in colloidal dispersions is essentially determined
by viscous effects. In physical terms the moving particle engenders in the
surrounding fluid a velocity field $\mathbf{v(r)}$, which produces entropy
according to the last term in (\ref{12}). Accordingly, most theoretical
works on colloidal thermophoresis rely on a hydrodynamic approach which
deals with the coupling of the mass flux $\mathbf{J}_{1}$ in (\ref{16}) and
the viscous pressure tensor (\ref{18}).

We consider colloidal particles dispersed in continuous solvent without
molecular structure. The chemical potential per particle $\mu =h-Ts$, or
partial free enthalpy, is given by the interaction enthalpy $h$ and the
translational entropy $s=-k_{B}\ln \phi _{1}$; the former is proportional to
the particle surface and the latter decreases with the particle content $%
\phi _{1}$. From the thermodynamic force (\ref{24}) one expects that $D$ and 
$D_{T}$ arise from the volume fraction and temperature derivatives,
respectively.\ 

In this Section we sketch the derivation of $D$ and $D_{T}$ in terms of
Stokes' equation $\eta \mathbf{\nabla }^{2}\mathbf{v=\nabla }P$, with the
solvent viscosity $\eta $, velocity field $\mathbf{v}$, and pressure $P$.\ A
rather simple physical picture emerges for the diffusion coefficient, where
the hydrodynamic flow corresponds to the Stokes drag of a particle subject
to an entropic force $-k_{B}T\nabla \phi _{1}$. Regarding the thermophoretic
mobility $D_{T}$, the relation between the thermodynamic force and the
particle velocity is less straightforward, but relies on an argument
developped by Derjaguin and on the concept of an effective slip velocity
close to a solid surface \cite{And89}. The latter provides a hydrodynamic
boundary condition with links viscous flux with the particle motion.\ 

\subsection{Stokes-Einstein diffusion coefficient}

Gradient diffusion in a collodial dispersion is determined by the interplay
between thermal noise and Stokes drag; in the present notation the
coefficient reads 
\begin{equation}
D=\frac{\phi _{1}\mu _{11}^{\phi }}{6\pi \eta R}=\frac{k_{B}T}{6\pi \eta R}.
\label{52}
\end{equation}%
The numerator results from the thermodynamic force $\nabla \mu $ exerted by
a concentration gradient, whereas the denominator accouts for Stokes
friction for a sphere of radius $R$. In the second equality we have used
that in a dilute suspension the thermodynamic factor simplifies according to 
$\phi _{1}\mu _{11}^{\phi }=k_{B}T$ \cite{deG84}.

In the absence of viscous effects and for $\phi _{1}\rightarrow 0$, Eq. (\ref%
{26a}) gives 
\begin{equation*}
D=B_{2}\phi _{1}\hat{\mu}_{11}^{\phi }=k_{B}TB_{2}/v_{1},
\end{equation*}%
with the solvent molecular mobility $B_{2}$ and the particle volume $v_{1}$ 
\cite{Wue14}. In the present macroscopic hydrodynamics approach (\ref{52}),
the mobility $B_{2}$ has disappeared, or rather is subsumed in the viscosity
parameter $\eta $. The dependence on the particle size $R$ is characteristic
for the solvent velocity field associated with the diffusing particle.

\subsection{Surface forces and slip velocity}

Now we turn to the thermophoretic mobility $D_{T}$, which describes
colloidal motion driven by a temperature gradient. For the case of thermal
diffusion, the coefficient (\ref{26b}) was obtained as the product of the
thermodynamic forces with the molecular mobilities. Thermophoresis is more
complex since the motion is related to a velocity field $\mathbf{v(r)}$ in
the surrounding fluid; the resulting viscous pressure $\Pi $ contributes
significantly to the entropy production.

As pointed out by Anderson \cite{And89}, the fundamental principle of
thermophoresis is similar to electrophoresis and to motion in concentration
gradients. As shown in Fig. 3, the temperature gradient induces a shear
stress within a boundary layer of thickness $\lambda $. At distances beyond $%
\lambda $, the resulting flow profile saturates at the effective slip
velocity $v_{S}$. Matching the far-field $\mathbf{v(r)}$ to the boundary
condition $v_{S}$, one finds that the particle moves at a velocity 
\begin{equation}
u=-D_{T}\nabla T=-\frac{2}{3}v_{S}.  \label{54}
\end{equation}%
Thus calculating $D_{T}$\ is reduced to the hydrodynamic problem of
evaluating the flow\ around the particle.

In the phenomenological equations, there are no cross-terms between the
particle current $\mathbf{J}_{k}$ and the viscous flux $\mathbf{\Pi }$. Yet
the hydrodynamic boundary conditions couple the particle motion to the fluid
velocity field. In the following we give Derjaguin's evaluation for this
coupling.

\begin{figure}[tbp]
\includegraphics[width=\columnwidth]{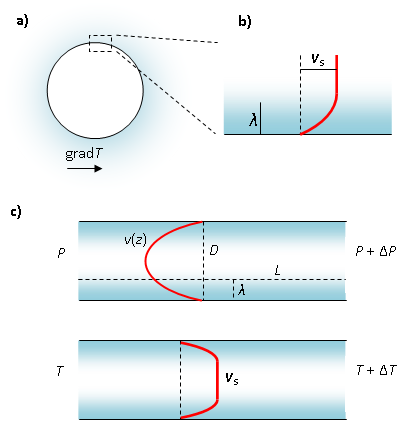}
\caption{a) Colloidal particle in a temperature gradient. The boundary layer
is characterized by the excess specific enthalpy $\hat{h}$ (grey-blue). b)
The case of a thin boundary layer ($\protect\lambda \ll R$) is equivalent to
a flat surface where, at distances beyond $\protect\lambda $, thermoosmosis\
leads to an effective slip velocity $v_{S}$. c). Heat and volume flows in a
capillary. The pressure gradient $\Delta P/L$ results in Poiseulle flow with
parabolic velocity profile $v(z)$. Since $\hat{h}=0$ in the core of the
capillary, the excess heat flow (\protect\ref{62}) occurs in the boundary
layers of thickness $\protect\lambda $. A temperature gradient $\Delta T/L$
results in thermal creep flow along the solid boundary; the velocity profile
in the boundary layer depends on the detail of $h(z)$.\ The constant
velocity in the core of the capillary can be determined from Onsager's
reciprocal law for the coefficients in (\protect\ref{56}). Then the
resulting slip velocity (\protect\ref{66}) applies equally well at the
surface of a colloidal particle and, finally, determines the thermophoretic
mobility according to (\protect\ref{54}). }
\end{figure}

\subsection{Onsager relation for enthalpy and volume flows}

In their 1941 paper on the thermoosmotic effect, Derjaguin and Sidorenkov
consider the conjugate heat flow instead of the thermally-driven velocity 
\cite{Der41,Der87}. In a second step, the latter is obtained from a
reciprocity relation. This is achieved in the geometry shown in Fig. 3c,
where the height $D$ of the capillary is much smaller than its length $L$
and its width $w$. The thermophoretic coefficient $D_{T}$ is obtained by
mapping the boundary problem on the surface of a colloidal particle to the
capillary flow velocity, according to (\ref{54}).

The thermodynamic forcers, that is the temperature gradient $\nabla T=\Delta
T/L$ and pressure gradient $\nabla P=\Delta P/L$, are constant and oriented
along the capillary axis. The corresponding fluxes, that is the flows of
heat and volume, are linear functions of the forces, 
\begin{subequations}
\label{56}
\begin{align}
J_{V}& =\frac{\dot{V}}{wD}=-L_{VQ}\frac{\nabla T}{T}-L_{VV}\nabla P, \\
J_{Q}& =\frac{\dot{Q}}{wD}=-L_{QQ}\frac{\nabla T}{T}-L_{QV}\nabla P.
\end{align}%
Here $\dot{V}$ and $\dot{Q}$ are the integrated volume and heat flows
through the capillary, whereas $J_{V}$ and $J_{Q}$ are the\ average current
densities, like $\mathbf{J}_{Q}$ in previous sections. The dimension of $%
\dot{Q}$ is energy/time, and that of $\dot{V}$ is volume/time.

The diagonal coefficients $L_{VV}$ and $L_{QQ}$ are described by the
Hagen-Poiseuille law for laminar flow and the thermal conduction of the
liquid, respectively. The off-diagonal coefficients account for the
cross-currents; $L_{QV}$ gives the pressure-driven heat flow, and $L_{VQ}$
the temperature-driven volume flow. \ According to Onsager's reciprocal
relations, these coefficients are identical, 
\end{subequations}
\begin{equation}
L_{QV}=L_{VQ}.  \label{58}
\end{equation}%
The transport coefficients are calculated from Stokes' equation $\eta \nabla
^{2}v=\nabla P$ for the velocity field $v(z)$, completed with the boundary
conditions $v(0)=0=v(D)$.

First consider the flows driven by a pressure difference. In a narrow
capillary, the velocity depends on the vertical coordinate $z$ only. Then
Stokes' equation reduces to $\eta \partial _{z}^{2}v=\nabla P$; it is solved
by 
\begin{equation}
v(z)=\frac{z(z-D)}{2\eta }\nabla P\ \ \ \ \ \ \ (0\leq z\leq D)  \label{60}
\end{equation}%
and results in the volume flow $\dot{V}=-(wD^{3}/12\eta )\nabla P$ and $%
L_{VV}=D^{2}/12\eta $. The heat flow consists of the excess enthalpy of the
liquid close to the upper and lower boundaries of the capillary, 
\begin{equation}
\dot{Q}=w\int_{0}^{D}dz\hat{h}(z)v(z).  \label{62}
\end{equation}%
The specific enthalpy $\hat{h}$ is measured with respect to that of the bulk
liquid, $\hat{h}_{\infty }$, such that $\hat{h}\rightarrow 0$\ outside the
interaction layers. Dividing the integral by $wD$ and identifiying with $%
J_{Q}=-L_{QV}\nabla P$, we find the transport coefficient%
\begin{equation}
L_{QV}=-\frac{1}{\eta }\int_{0}^{D/2}dzz\hat{h}(z).  \label{64}
\end{equation}%
Here we have used that the lower and upper boundaries of the capillary carry
identical amounts of heat, and have inserting the velocity profile close to
the lower boundary, $v(z)=-zD\nabla P/2\eta $ for $z\ll D$.

Now we turn to the temperature-driven volume flow at constant pressure.
Except for the boundary layer, the velocity profile is constant across the
capillary; to leading order in $\lambda /D$, the volume current reads $%
J_{V}=v_{S}$. Identifying this with $J_{V}=-L_{VQ}\nabla T/T$ and using the
reciprocal relation (\ref{58}), we find 
\begin{equation}
v_{S}=-\frac{\nabla T}{\eta T}\int_{0}^{\infty }dzz\hat{h}(z).  \label{66}
\end{equation}%
We have used that $\hat{h}$ is finite in the boundary layer of thickness $%
\lambda \ll D$ only and vanishes in the core of the capillary; thus we have
replaced the upper bound $D/2$ with infinity. (In passing we note that the
additional factor 2 occuring in previous work \cite{Der87,Pia04} is related
to the missing $\frac{1}{2}$ in the Poiseuille velocity profile used there.)

\subsection{Thermophoretic and Dufour coefficients}

The above cross-coefficient $L_{QV}$ applies to any surface with an excess
enthalpy $\hat{h}$ within a thin boundary layer. In particular, $v_{S}$
describes the quasislip velocity $v_{S}$ occuring close to the surface of a
colloidal particle, as illustrated in Fig. 3b. From the hydrodynamic
boundary condition (\ref{54}) one readily obtains the thermophoretic
mobility of the particle, 
\begin{equation}
D_{T}=-\frac{2}{3\eta T}\int_{0}^{\infty }dzz\hat{h}(z)=\frac{2}{3}\frac{%
L_{QV}}{T}.  \label{69}
\end{equation}%
In the second equality we have used (\ref{64}) with $\lambda /D\rightarrow 0$%
. Note that $D_{T}$ does not depend on the particle radius $R$. (This is
valid as long as the particle radius is larger than the interaction length, $%
R\gg \lambda $. In the opposite case, the numerical prefactor $\frac{2}{3}$
has to be replaced with unity \cite{Wue10}.)

Thus thermophoresis stems from the volume flow $\dot{V}$ driven by
thermo-osmosis along a solid-liquid interface. The coefficient $%
L_{VQ}=L_{QV} $ has been evaluated through the conjugate phenomenon, that
is, the heat flow $\dot{Q}$ due to a Poiseuille flow. The second equality in
Eq. (\ref{69}) relies explicitly on the Onsager reciprocity law (\ref{58}).
It is clear that $D_{T}$ as defined above, has no relation with the Dufour
effect $D_{F}$. In physical terms, the above argument does not lead to heat
flow induced by a non-uniform colloidal volume fraction, and the actual
conjugate flux $\dot{Q}$ has nothing do with a non-uniform composition.

Still, in a colloidal suspension with non-uniform volume fraction $\phi $,
there may be an additional heat flux $\dot{Q}_{F}\propto L_{Q1}^{\prime
}\nabla \phi $, and the reciprocal coefficient $L_{1Q}^{\prime }$ is
expected to contribute an additional term to (\ref{69}). The many data
collected for colloidal thermophoresis suggest, however, that this extra
term is small. In particular, experiments show that $D_{T}$ is independent
of the particle size; it would be rather surprising if the heat flow $\dot{Q}%
_{F}$, and thus the thermal diffusion and Dufour coefficient $L_{Q1}^{\prime
}=L_{1Q}^{\prime }$, did not vary with the particle size.

\section{Discussion}

\subsection{Are $D_{T}$ and $D_{F}$ reciprocal coefficients?}

Reciprocicity is an undoubted property of the cross-coefficients in
Onsager's equations, such as $L_{1Q}^{\prime }$ and $L_{Q1}^{\prime }$ in (%
\ref{16}), or $L_{QV}$ and $L_{VQ}$ in (\ref{58}). This paper addressed the
question to what extent these reciprocal relations imply $D_{T}=D_{F}/T$ for
the thermal diffusion and Dufour coefficients \ in liquids, and under which
conditions the identification of Eq. (\ref{20}) is valid. In the absence of
scalar and vector fluxes, we find that the model equations (\ref{2}) are
identical to Onsager's linear relation (\ref{16}), implying $D_{T}=D_{F}/T$.

Adding a chemical reaction and coupling to the vector fluxes through (\ref%
{38}), we obtain a finite $D_{T}$ even for $L_{1Q}^{\prime }=0$. The
corresponding micelle current (\ref{40}) is driven by the equilibrium
composition of the chemical reaction. Thus we find that Eq. (\ref{20}) is
not satisfied in the presence of chemical reactions, in agreement with the
more formal study by de Groot and Mazur \cite{deG84}.

A similar result is obtained for viscous effects. Thermophoresis in
colloidal suspensions is governed by hydrodynamic flow in the vicinity of
the particle, as shown in general by Anderson \cite{And89} and worked out in
detail for electric-double layer interactions \cite{Wue10,Esl14}. The
mobility $D_{T}$ is given by the interaction enthalpy in the boundary layer.
The corresponding Onsager coefficient $L_{QV}$ is not related to Dufour
effect and its coefficient $L_{Q1}^{\prime }$.

Our analysis suggests that the mobility $D_{T}$ defined in (\ref{2}) and
measured in many experiments, comprises different contributions: thermal
diffusion, thermophoresis, spatially varying chemical reaction rates,... \
In each case, there is a conjugate effect with a reciprocal coefficient.
Above we have discussed the relation between thermal diffusion and the
Dufour effect ($L_{1Q}^{\prime }=L_{Q1}^{\prime })$, and thermophoresis and
the pressure-driven heat flow ($L_{VQ}=L_{QV}$). Similarly, the equilibrium
between the molecular and micellar states in a surfactant solution, feeds a
steady micelle current (\ref{40}) that is proportional to the temperature
gradient; the coeffiicient $D_{T}$ is determined by the reaction rates of
micelle formation.

\subsection{Thermal diffusion or thermophoresis?}

Most authors use \textquotedblleft thermal diffusion\textquotedblright\ for
molecular mixtures and \textquotedblleft thermophoresis\textquotedblright\
for colloidal suspensions, though in both cases the coefficient $D_{T}$ is
defined by (\ref{2}). The above discussion gives a more precise meaning to
this disctinction. In the case of thermal diffusion, viscous effects are
absent, and $D_{T}$ and the Dufour coefficient are related by reciprocity.
In the stationary state the particle currents vanish, $\mathbf{J}_{1}=0$,
and so does the corresponding entropy production.

Thermophoresis, on the other hand, is determined by viscous flow. Then the
mobility $D_{T}$ is given by the Onsager cross-coefficient $L_{VQ}$ which
describes volume flow due to a temperature gradient. As worked out in a
previous paper \cite{Wue13}, the entropy production related to the particle
fluxes vanishes in the steady state $\mathbf{J}_{1}=0$, the viscous flux
continues to dissipate energy. The hydrodynamic flow around each particle
maintains a finite rate of entropy production $\mathbf{\Pi :G}>0$.

\subsection{Comparison with experiment and simulations}

According to the preceding discussion, the reciprocity relation between $%
D_{T}$ and $D_{F}$ is valid for thermal diffusion only. In other words,
reciprocity requires that diffusion and thermal diffusion coefficients can
be expressed through the mobilities $B_{k}$ as in (\ref{26}). In view of the
fit of \ $S_{T}=D_{T}/D$ in Fig.\ 1, the alkane data\ are not in
contradicton with the thermal diffusion picture. In order to obtain
conclusive evidence, one would have to consider the coefficients $D_{T}$ and 
$D$ separately, and this at different composition. \ On the contrary, data
on polymer solutions and particle suspensions rather agree with
thermophoresis mechanism, indicating that viscous effects prevail if the
solute is much larger than the solvent molecules \cite{Gid76,Bro81}. Yet the
break-down of the thermal diffusion picture may occur well before viscous
behavior in the sense of macroscopic hydrodynamics sets in.

In technical terms, (\ref{26}) ceases to be valid if the molecules respond
differently to the thermal and concentration gradients in the thermodynamic
force (\ref{19}). Yet nothing is known about the underlying mechanisms and
the relevant parameters. At present it is not even clear whether the thermal
diffusion picture applies to mixtures of organic molecules of similar size
and weight, such as benzene, cyclohexane, and alkanes. Whereas thermal
diffusion data are available for many systems \cite%
{Har13,Deb01,Kit04,Kit05,Har06}, there are only few studies on the Dufour
effect \cite{Ras70,Hor78}. Comparison of measured $D_{T}$ and $D$ with (\ref%
{26}) could provide valuable information.

In recent years, molecular dynamics simulations have become a powerful tool
for studying the thermal diffusion and transport properties. Besides the
dependencies on molecular enthalpy, mass, and size \cite{Mue99,Art07,Gal08},
both diagonal and off-diagonal Onsager coefficients have been evaluated \cite%
{Mac86,Pao87,Mil13,Arm14}. For the investigated Lennard-Jones systems, the
simulations confirm reciprocity of the thermal diffusion and Dufour
coefficients. So far there is no systematic numerical study of the validity
of (\ref{4}) and (\ref{26}). The above discussion suggests a break-down of
the thermal diffusion upon differentiating the two components in molecular
size or shape.

\section{Summary}

In this paper we discussed the coefficient $D_{T}$ of thermally driven
transport in three situations, where the entropy production is dominated by
diffusion, chemical reactions, or viscous flow. In the first case, $D_{T}$
is determined by the molecular enthalpy and diffusion constants $B_{i}$
according to (\ref{26}).\ The conjugate effect is the Dufour effect, and $%
D_{T}$ and $D_{F}$\ are reciprocal coefficients.

As an example of chemical reactions, the temperature-dependent equilibrium
between micellar and molecular states in a surfactant solution, imposes a
steady-state diffusion current. The resulting coefficient $D_{T}$ in (\ref%
{40}) is given by reaction rates and the micelle enthalpy $H$. As the
reciprocal effect, an externally imposed micelle current feeds a stationary
reaction flux $\mathcal{J}=\dot{\phi}_{1}$.

\begin{figure}[tbp]
\includegraphics[width=\columnwidth]{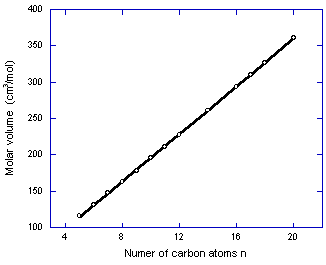}
\caption{Molar volume of normal alkanes. The circles give the measured
values at room temperature, the solid line has been calculated from Eq. (%
\protect\ref{72b}). }
\end{figure}

If dissipation is dominated by viscous stress, as for colloidal suspensions
in (\ref{69}), $D_{T}$ depends on the interaction enthalpy and the solute
viscosity. The reciprocal cross-coefficient $L_{QV}$ accounts for advective
heat flow in a pressure gradient.

Thus in the presence of chemical reactions or viscous effects, $D_{T}$ is
not related to the Dufour effect. In future work, it would be interesting to
study whether and to which extent, the thermal diffusion picture (\ref{26})
is valid in binary molecular mixtures.

\textbf{Acknowledgement. }This work was supported by Agence Nationale de la
Recherche through contract ANR-13-IS04-0003.

\section{Appendix: Alkane enthalpy and volume}

The fit of alkane Soret data in Fig. 1 relies heavily on the variation of
the molecular enthalpy and volume (\ref{72}) with the number $n$ of carbon
atoms. In particular, the relative off-set values of $0.65$ and $2.02$
determine to a large extent the theoretical curve. If these numbers were
identical, the specific enthalpy $\hat{h}=h/v$ would be independent of $n$,
and $S_{T}$ would vanish for all mixtures; exchaning the off-set values
would result in a positive Soret effect for the lighter component.

An at least qualitative explanation for the off-set of $0.65\times
4.64\approx 2.9$ kJ/mol is given by a simple geometrical argument for
polymers on a 3D cubic lattice. A monomer has 6 couplings with next
nearest-neighbor molecules and a dimer has $\frac{10}{2}=5$ such couplings
per unit. For a rigid high polymer there are $4(n+2)$ couplings, that is,
about 4 per monomer, and a slightly smaller value occurs for flexible
polymers. These numbers are very close to the factor in (\ref{72a}).

The volume off-set accounts for the fact that the density of shorter alkanes
is smaller; it is related to the larger entropy of short chains. Fig. 4
shows measured values of the molecular volume $v_{n}$.\ The solid curve,
calculated from (\ref{72b}), provides a very good fit to these data.

\end{document}